# PREDIS-MHI Thermal Data: Thermal readings from a section of a sub-metered tertiary multi-use building

**Seun OSONUGA[1†], Ali CHOUMAN[1,2], Muhammad-Salman SHAHID[1], Benoit DELINCHANT[1], Frederic WURTZ[1]**

**[1] Univ. Grenoble Alpes, CNRS, Grenoble INP\*, G2Elab, 38000 Grenoble, France**

**[2] Centre Scientifique et Technique du Batiment (CSTB), F-06904 Sophia Antipolis, France**

**\* Institute of Engineering Univ. Grenoble Alpes**

**[†]seun.osonuga@g2elab.grenoble-inp.fr**

ABSTRACT. Tertiary buildings could be an important lever to meet the goals necessitated by the energy transition. The availability of high-quality datasets from this sector will be a crucial enabler in meeting these goals by developing and testing new energy management approaches in the buildings. In this paper, we present the thermal energy datasets available and published online for the PREDIS-MHI zone of the GreEn-ER building, a tertiary building with more than a thousand sensors used for research, teaching, and administrative activities in Grenoble. PREDIS-MHI platform is a net-zero sub-section that is energetically isolated from the rest of the building. Its data has been used in a wide range of applications from indoor temperature forecasting, thermal simulation calibration, and even occupant comfort experiments.

## 1. INTRODUCTION

Although not the majority of the world's building stock, tertiary multi-use buildings contribute a significant proportion to the greenhouse gas (GHG) emissions from buildings in most of the world's cities. In France, these buildings represent about 24% (973.4 million $m^2$) of the total floor area, and account for about a third of the final energy consumption in buildings. Most of the energy consumed in building is on spent on heating and cooling. The French Construction and Housing Code had set an obligation for tertiary buildings to reduce their final energy consumption by 40% in 2030, 50% in 2040, and 60% in 2050. According to ADEME, tertiary buildings in France account for almost a third of the energy consumed in the built environment.

As such this article presents the PREDIS-MHI Thermal Data, a thermal energy dataset from a thermally and electrically separated section of the GreEn-ER building in Grenoble (Osonuga et al. 2024). In this article, the HVAC system for the PREDIS-MHI platform is described as well as its machine-readable corollary which is based on the BRICK schema (Fierro et al. 2019). Finally, several use cases are presented of the dataset with special mention given to the use case of an energy simulator calibration.

## 2. DESCRIPTION OF THE PREDIS-MHI PLATFORM

The GreEn-ER building, in which the PREDIS-MHI platform is found, is a five-storey mixed-used tertiary building in Grenoble France with a total floor space of about 22,700 m$^2$ (Delinchant et al. 2016). It houses about 2000 researchers and students who frequent the building either for the resident laboratory G2Elab or the engineering school, ENSE3.

The PREDIS-MHI platform, with a floor space of about 600 m$^2$, is located mostly in the corner of the fourth floor of the GreEn-ER building as seen in Figure 1. There are also a few other rooms on the second floor of the same building. The shown area is made of a classroom (Room 4A020), an exposition room (Room 4A013), and 6 offices (Rooms 4A014 – 4A019) as shown on the left side of Figure 1. More details about the dimensions of the PREDIS-MHI platform and the larger building can be found in the 3d model of the platform that is part of the PREDIS-MHI dataset (Osonuga et al. 2024).

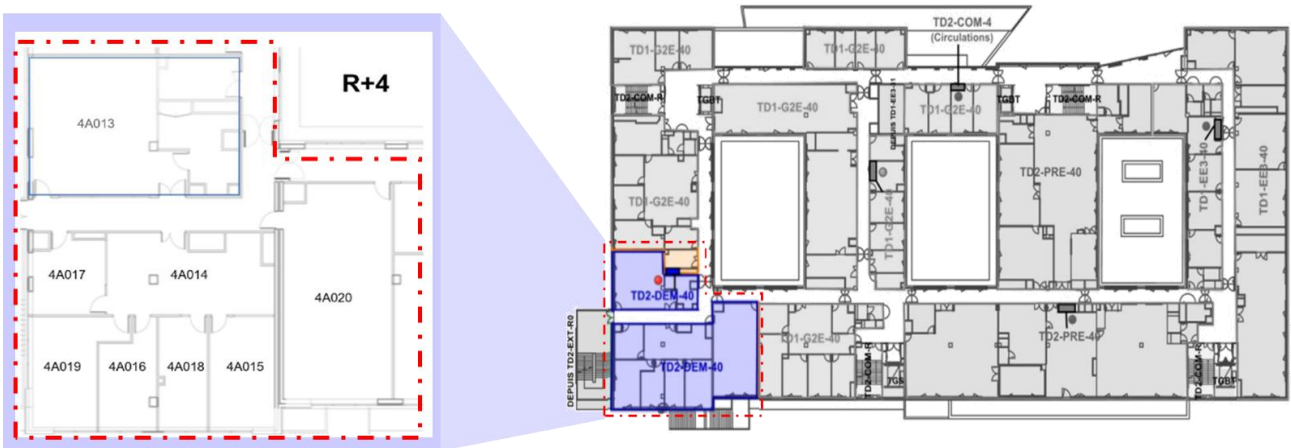

*Figure 1: The PREDIS-MHI Platform in the GreEn-ER Building*

The PREDIS-MHI platform is served by an exclusive HVAC system. The system has a dedicated Air Handling Unit (AHU) located on the building's rooftop whose schematic is shown in Figure 2. Hydroponic sources from the local district heating network provide the heating and cooling of the HVAC system. Each room is treated as an individual HVAC zone with the ability to regulate the set points for the temperature and $CO_2$ levels. The temperature control is mostly provided in two modes - through reheat VAV coils, or the installed radiant ceilings. Each of the rooms can be operated in the one of the above modes or in a third that only dies ventilation without heating or cooling. In this third mode, the blown air will be the same temperature as the air at the AHU which is preconditioned to 21 degrees in the summer months and 19 degrees in the winter months. A machine-readable description of this system is available as part of the dataset.

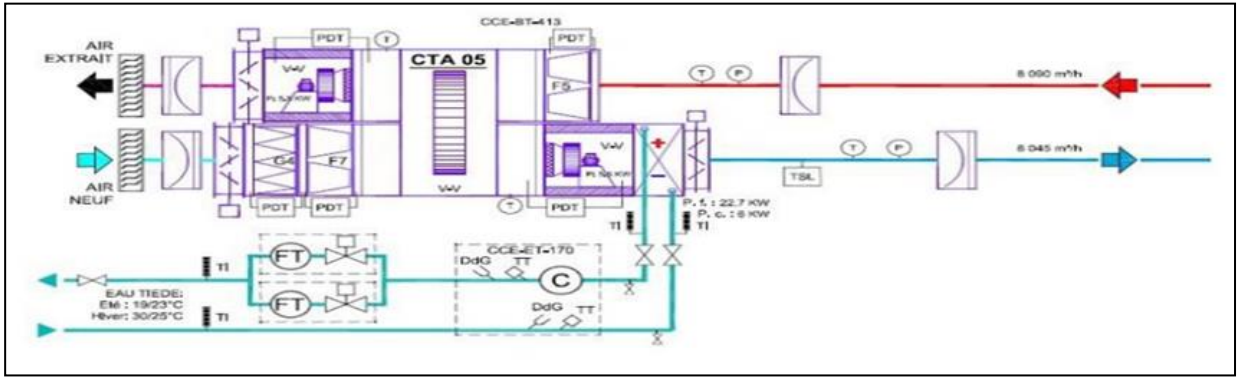

*Figure 2: The Air Handling Unit*

# 3. PREDIS-MHI DATA

## 3.1. DESCRIPTION OF THE DATA

The dataset includes the unsampled data from the building management systems as well as building meta-data. Figure 3 shows the arrangement of the dataset with room-level data including temperature, $CO_2$, and Airflow rates. There is also specific data from the Air handling unit and general data in the root folder of the dataset.

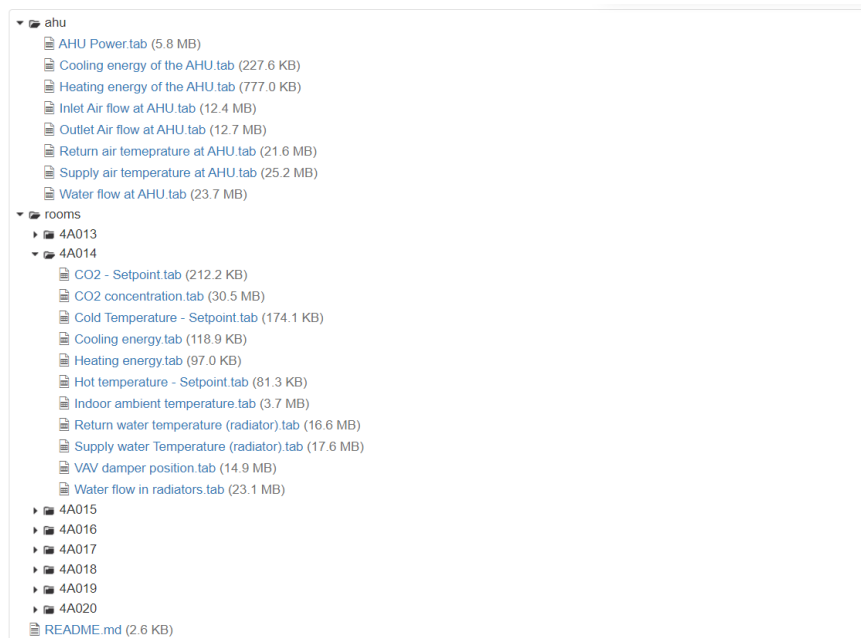

*Figure 3: Folder structure of the PREDIS-MHI dataset on recherche.data.gouv*

## 3.2. DATASET CONSTRUCTION

The GreEn-ER building is equipped with numerous sensors and actuators which measure the building conditions in real-time and enable the modification of the operating conditions from the building management system, known as EcoStruxure (Mora et al. 2012). The data storage process comprises multiple steps. The data is measured by installed sensors in real time and is stored in the memory of the concerned programmable logic controller (PLC). The PLC has a data retention of 15 days; therefore, every new registered value pushes out the oldest registered value from the memory. To avoid the loss of this data, the values from these PLCs is copied on the SQL database. This SQL database does not have a retention policy. No pre-treatment of data is realised at this stage. This data can be visualized by using the EcoStruxure web interface. Only a limited part of this data is related to the demo zone of the GreEn-ER building, therefore, the data of the demo zone is at disposal for research and teaching. If necessary, a pre-treatment of data is realised at this stage. This data is retrievable from the time-series database on the MHI server (known as influxDB) and can be visualized publicly using the Grafana web interface tool. Figure 4 illustrates a simplified version of the registration and visualisation process of GreEn-ER data.

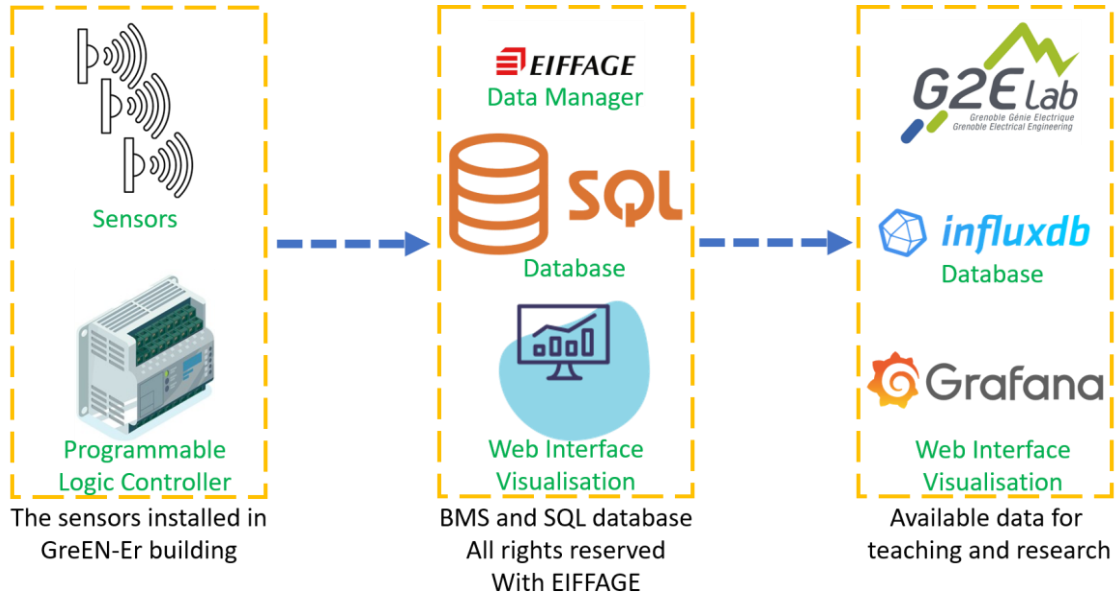

*Figure 4: Simplified schematic diagram of GreEn-ER data registration and visualisation process*

## 3.3. VISUALIZATION OF DATA

The raw data collected was a set of DataFrame files containing the values recorded by each sensor for a standard time step. To use and visualize the information given by the sensors, a data treatment script was developed. This script generates graphs of both overall and specific heating and cooling consumption, showcasing energy trends throughout a selected year. It also incorporates indoor temperature and airflow information into the analysis. Data of various measures are consolidated into a single DataFrame, with values standardized to a one-hour interval through interpolation. For temperature datasets containing missing values, these gaps are filled using data from the previous time step. A graphical user interface within the script allows users to select specific datasets for visualization, presenting annual trends of data. Graphs illustrate the operational patterns of the ventilation system, which is inactive overnight as shown in Figure 5, where the air change rate (ACR) value , indicating the volume of fresh air introduced into the space per hour, exhibits elevated levels from 6 AM to 6 PM during the daytime and significantly decreases during the night-time.  Similarly, temperature variations within the rooms align with predetermined economy and comfort thresholds for night and day periods, respectively.

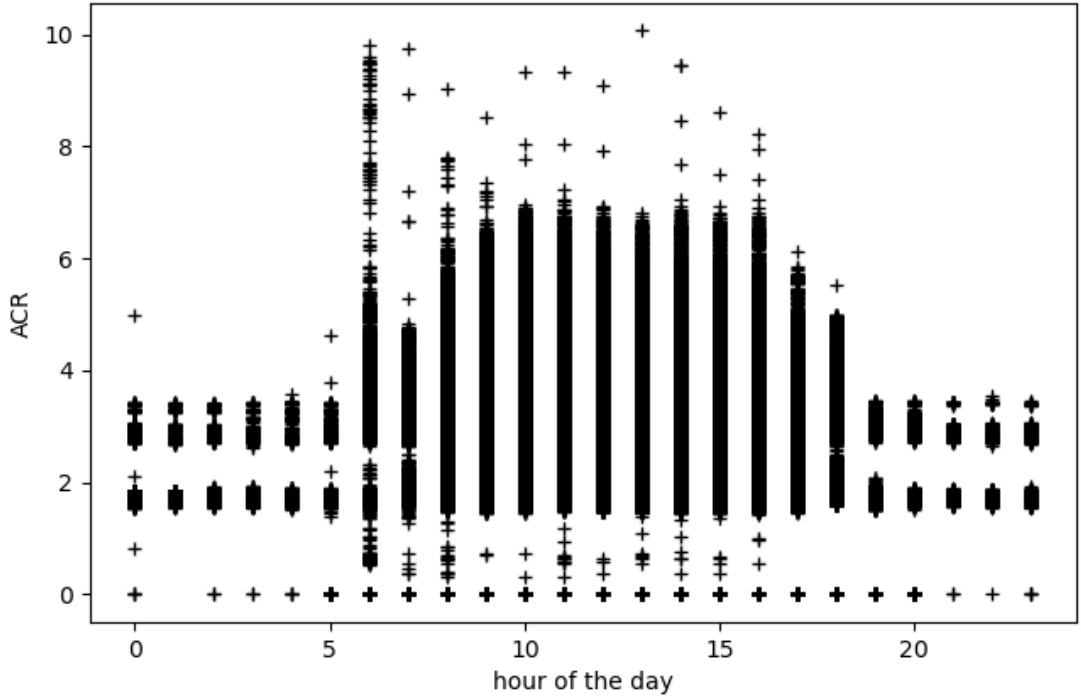

*Figure 5: The evolution of air change rate in the rooms according to the day hour.*

The energy consumption graphs indicate heating from January to April and October to December, with cooling occurring from July to September (see Figure 6).

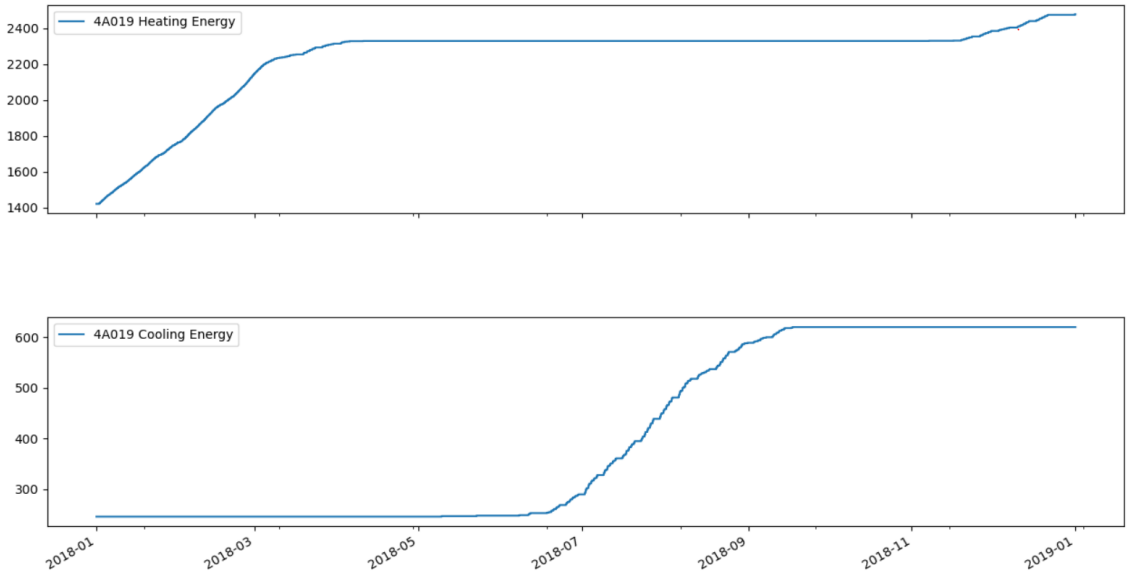

*Figure 6: Example of heating and cooling energy in the room 4A019 of PREDIS-MHI zone.*

## 4. VALUE OF THE PREDIS-MHI THERMAL DATASET

The dataset has been in use not only for teaching at ENSE3 and research in G2ELab, rather it is also used for other purposes. The building took part in the energy flexibility competition during the academic year 2021-2022, launched by RTE (the French transmission operator) and a private entity, known as A4MT (short for action for market transformation). The energy savings obtained by modifying the thermal setpoints is calculated thanks to this data.

In addition to this, the G2ELab is part of an ongoing experiment in its living lab. This experiment is part of a research work carried out in the framework of a European project, known as COLLECTiEF (Aghaei et al. 2023). The purpose of this project is to play on the energy flexibility capacity of the HVAC system while considering the human comfort. The analysis of data is ongoing, whereas the initial results are published regarding the indoor thermal comfort analysis (Papadopoulos et al. 2023). Based on these studies, it is envisaged that an automated and robust machine learning based mechanism will be applied to achieve optimum energy flexibility without compromising human comfort in the building.

## 4.1. USE OF THE DATA

A practical application of PREDIS-MHI data use involves developing a thermal model through the Dimosim simulation tool. This model serves as a foundation for testing and evaluating advanced algorithms for managing energy systems. Dimosim is a bottom-up dynamic simulation platform, developed by the CSTB (Garreau et al. 2021). This tool is based on an object-oriented structure. The models of buildings and energy components are based on a physical description.

To run the Dimosim model, several pieces of information were needed. First, a detailed weather data file in .epw format is required to provide weather conditions. This is sourced from the weather station on the building (PREDIS-MHI (GreEn-ER) 2015). Additionally, the simulation needs to specify the duration of each timestep. A Geojson file describing the building is also crucial. This file includes details about the building's shape and design, in addition to the data of the physical characteristics in the simulated zone. Other information such as thermal energy consumption and energy system specifications are also needed.

The building's technical documentation is used to extract the essential data for the Geojson file, including the building's footprint and the thermal properties of its walls and windows. In addition, a meteorological file is created using data from the building's meteorological station. Furthermore, it's important to estimate both the total energy used to heat and cool the lab and the individual energy used by each office. For this purpose, a large dataset was collected from the lab's database, recording energy and temperature sensor measurements from June 2015 until 2022. For each room within the PREDIS-MHI zone, a virtual global energy meter is designated to approximate the aggregate thermal energy supplied by various energy systems. To enhance accuracy in this simulation, the analysis did not rely on readings from these virtual meters. Instead, calculations of the actual energy distributed to each room were conducted using data from real energy meters. Each office is conditioned using two methods: radiant ceilings and ventilation. The radiant ceilings are equipped with cumulative energy meters, which can calculate the ceiling's thermal energy consumption over time. In the case of ventilation, the energy output of the ventilation air flow is determined by two types of sensors. The first type is a cumulative energy meter at the air handling unit (AHU) level, which allows the estimation of the total energy produced and distributed by the AHU. The second type includes zonal airflow meters. This data is analyzed to determine the proportion of air delivered to the offices as a percentage of the total air delivered from the AHU. The energy delivered to each office is then estimated by multiplying this calculated airflow percentage by the total energy delivered by the AHU.

The collected raw data comprised DataFrame files, which documented the measurements from each sensor at consistent time intervals. A data processing script was crafted to transform and graphically represent the data, enabling the visualization of both overall and specific heating and cooling energy consumption, in addition to the air flow data. The external conditions data are also captured and treated to create the meteo file associated with the model. Indoor temperature and airflow data within the rooms were recorded every 10 minutes, although some datasets contained missing values, which were filled using the data from the preceding time step. Additionally, external weather conditions, including temperature and both direct and diffused solar radiation, were tracked with the same 10-minute resolution. For energy consumption metrics, cumulative energy meters were used, which log data each time a 1 kWh threshold is exceeded, rather than at fixed intervals. To align with the hourly time step of the Dimosim simulation, all collected data was standardized to a one-hour resolution by interpolation. The 3D representation of the model in Figure 7 shows three thermal zones-Greener 0, Greener 1, and Greener 2-which serve as boundary conditions for the simulated PREDIS-MHI area, referred to as G2ELab.

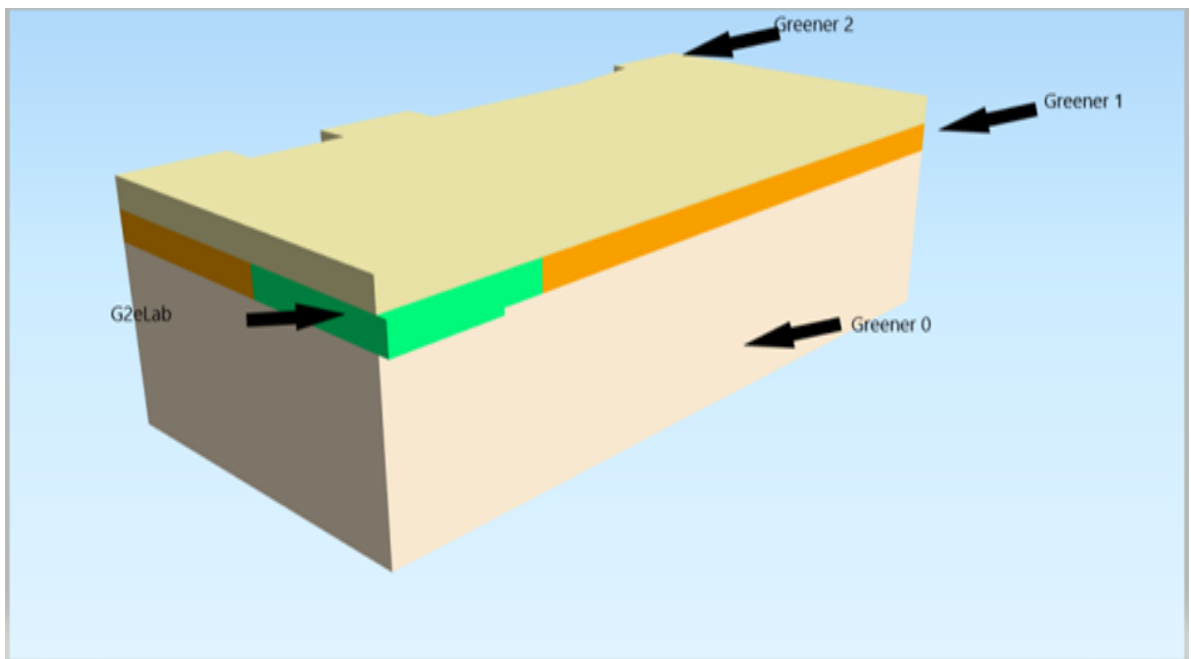

*Figure 7: 3d Q-GIS view of the model used in Dimosim simulation.*

## 5. CONCLUSIONS

This article has introduced the PREDIS-MHI Thermal a thermal energy dataset from a thermally and electrically separated section of the GreEn-ER building in Grenoble. In addition to this, the lifecycle was used to describe the process for the available energy datasets from the GreEn-ER building, a mixed-use education building in France.

As future work, we intend to publish the rest of the datasets from the GreEn-ER building. These include the thermal data from the rest of the building, the electrical data from the PREDIS-MHI platform which contains a solar PV array and EV charging points as well as the weather data from the building's weather station.